\documentclass[twocolumn]{aastex63}

\usepackage{amsmath}
\usepackage{gensymb}
\usepackage{booktabs}
\usepackage{xcolor}

\newcommand{\msun}{${\rm M_{\sun}}$}

\def\ltsima{$\; \buildrel < \over \sim \;$}
\def\simlt{\lower.5ex\hbox{\ltsima}}
\def\gtsima{$\; \buildrel > \over \sim \;$}
\def\simgt{\lower.5ex\hbox{\gtsima}}
\def\kms{{\rm\,km\,s^{-1}}}

\def\kpc{{\rm\,kpc}}

\def\msun{{\rm\,M_\odot}}

\def\g{{\rm\,g}}

\def \r {r$^{1/4}$ }


\def\deg{^\circ}

\def\s{\ifmmode \widetilde \else \~\fi}
\def\={\overline}

\def\spose#1{\hbox to 0pt{#1\hss}}

\def\lta{\mathrel{\spose{\lower 3pt\hbox{$\mathchar"218$}}
     \raise 2.0pt\hbox{$\mathchar"13C$}}}
\def\gta{\mathrel{\spose{\lower 3pt\hbox{$\mathchar"218$}}
     \raise 2.0pt\hbox{$\mathchar"13E$}}}
\def\Dt{\spose{\raise 1.5ex\hbox{\hskip3pt$\mathchar"201$}}}    
\def\dt{\spose{\raise 1.0ex\hbox{\hskip2pt$\mathchar"201$}}}    

\def\r{${\rm r^{1/4}}$}

\def\dotsfill{\leaders\hbox to 1em{\hss.\hss}\hfill}
\def\sun{\odot}
\def\earth{\oplus}

\def\Gyr{{\rm\,Gyr}}

\def\ltsima{$\; \buildrel < \over \sim \;$}
\def\gtsima{$\; \buildrel > \over \sim \;$}
\def\lsim{\lower.5ex\hbox{\ltsima}}
\def\gsim{\lower.5ex\hbox{\gtsima}}
\def\lapp{\ifmmode\stackrel{<}{_{\sim}}\else$\stackrel{<}{_{\sim}}$\fi}
\def\gapp{\ifmmode\stackrel{>}{_{\sim}}\else$\stackrel{<}{_{\sim}}$\fi}

\defcitealias{V21}{V21}
\usepackage{subfigure}

\submitjournal{ApJL}
\received{June 21, 2022}
\accepted{August 02, 2022}

\shorttitle{The Typhon polar stream}
\shortauthors{Tenachi et al.}
\graphicspath{{./}{figures/}}

\begin{document}

\title{Typhon: a polar stream from the outer halo raining through the Solar neighborhood}

\correspondingauthor{Wassim Tenachi}
\email{wassim.tenachi@astro.unistra.fr}

\author[0000-0001-8392-3836]{Wassim Tenachi}
\affiliation{Universit\'e de Strasbourg, CNRS, Observatoire astronomique de Strasbourg, UMR 7550, F-67000 Strasbourg, France}

\author{Pierre-Antoine Oria}
\affiliation{Universit\'e de Strasbourg, CNRS, Observatoire astronomique de Strasbourg, UMR 7550, F-67000 Strasbourg, France}

\author[0000-0002-3292-9709]{Rodrigo Ibata}
\affiliation{Universit\'e de Strasbourg, CNRS, Observatoire astronomique de Strasbourg, UMR 7550, F-67000 Strasbourg, France}

\author[0000-0003-3180-9825]{Benoit Famaey}
\affiliation{Universit\'e de Strasbourg, CNRS, Observatoire astronomique de Strasbourg, UMR 7550, F-67000 Strasbourg, France}

\author[0000-0002-8129-5415]{Zhen Yuan}
\affiliation{Universit\'e de Strasbourg, CNRS, Observatoire astronomique de Strasbourg, UMR 7550, F-67000 Strasbourg, France}

\author[0000-0002-0544-2217]{Anke Arentsen}
\affiliation{Universit\'e de Strasbourg, CNRS, Observatoire astronomique de Strasbourg, UMR 7550, F-67000 Strasbourg, France}

\author[0000-0002-1349-202X]{Nicolas Martin}
\affiliation{Universit\'e de Strasbourg, CNRS, Observatoire astronomique de Strasbourg, UMR 7550, F-67000 Strasbourg, France}

\author[0000-0002-7507-5985]{Akshara Viswanathan}
\affiliation{Kapteyn Astronomical Institute, University of Groningen, Postbus 800, 9700 AV, Groningen, the Netherlands}


\begin{abstract}
We report on the discovery in the Gaia DR3 astrometric and spectroscopic catalog of a new polar stream that is found as an over-density in action space. This structure is unique as it has an extremely large apocenter distance, reaching beyond 100 kpc, and yet is detected as a coherent moving structure in the Solar neighborhood with a width of $\sim 4\kpc$. A sub-sample of these stars that was fortuitously observed by LAMOST has a mean spectroscopic metallicity of $\langle {\rm [Fe/H]}\rangle = -1.60^{+0.15}_{-0.16}$~dex and possesses a resolved metallicity dispersion of $\sigma({\rm [Fe/H]}) = 0.32^{+0.17}_{-0.06}$~dex. The physical width of the stream, the metallicity dispersion and the vertical action spread indicate that the progenitor was a dwarf galaxy. The existence of such a coherent and highly radial structure at their pericenters in the vicinity of the Sun suggests that many other dwarf galaxy fragments may be lurking in the outer halo.
\end{abstract}


\section{Introduction}
\label{sec:Introduction}

One of the principal goals of the Gaia space mission \citep{GaiaMission} is to survey the Milky Way, so as to allow us to understand how our home galaxy was built up over cosmic time. Although we only observe the end state of this majestic structure, fortunately the processes of formation and growth have left copious amounts of evidence in the form of debris that is now scattered throughout our Galaxy \citep{2006ApJ...642L.137B,2018ApJ...862..114S,2021ApJ...914..123I,Malhan2022}. Some of these residues are due to the accretion of small galaxies and globular clusters, which disrupted under the action of tidal forces, leaving long stellar streams. In some cases they can still remain as elongated structures, many billion years after the dissolution of their progenitors \citep{Helmi2008}. Studying these structures is of great importance since their trajectories probe the galactic acceleration field and the underlying dark matter distribution \citep[e.g.,][]{2010ApJ...712..260K,2013MNRAS.433.1826S,Malhan2019,Ibata2021}.

A particularly powerful means to uncover such fossil remnants is by searching for groups of stars with common integrals of motion. Action coordinates are perhaps the best choice for this, as they are adiabatic invariants that will have been preserved along orbits if the Milky Way's potential evolved only slowly through time \citep{BinneyTremaine}. However, to transform our stellar measurements into actions (and their conjugate angles), we require the full six-dimensional positions and velocities. With present instrumentation this is only achievable close to the Solar position in the Galaxy.

\begin{figure*}[!htb]
\centering
  \includegraphics[angle=0,  clip, width=\textwidth]{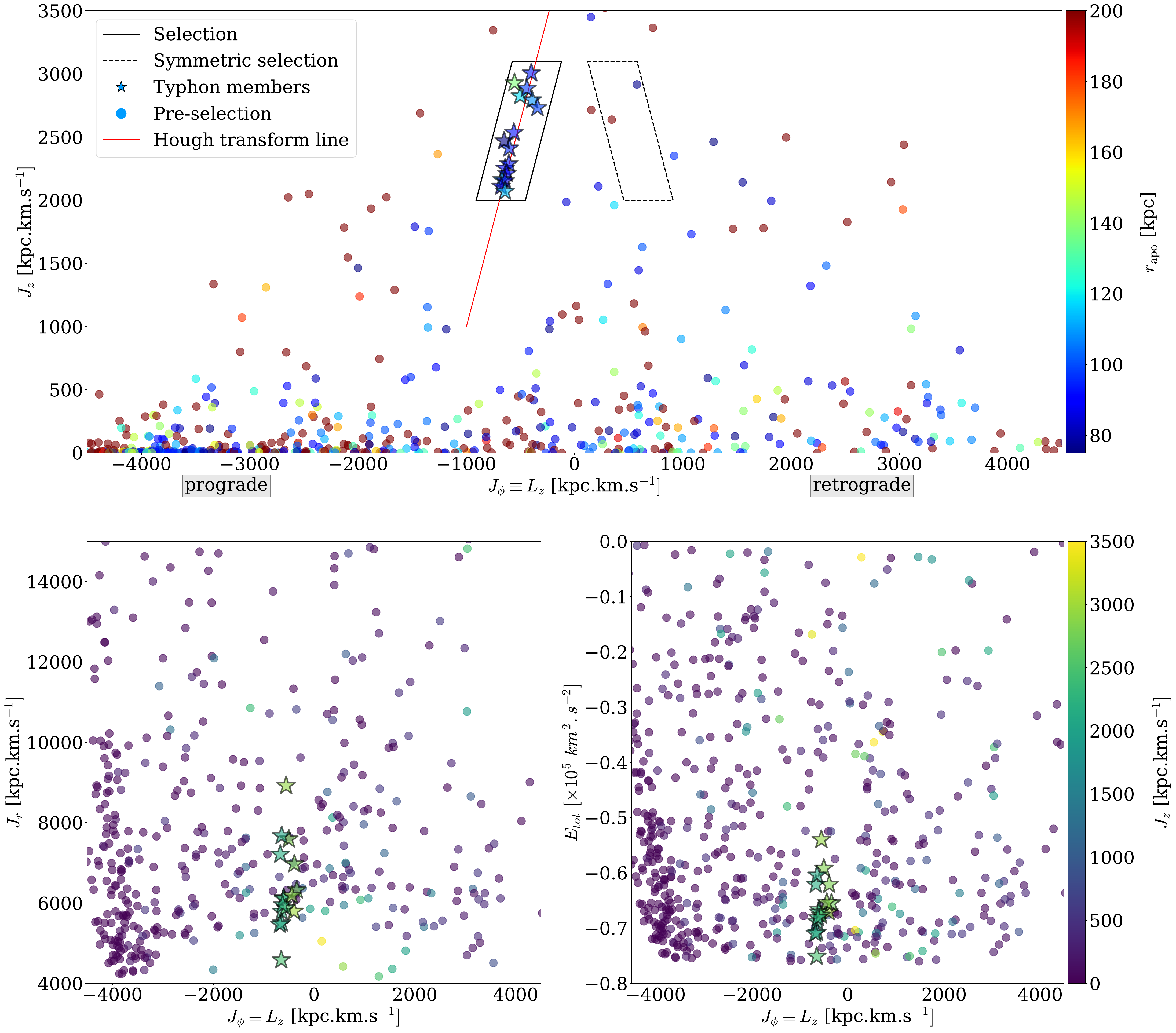}
   \caption{Actions and total energy of Typhon members (indicated by star-symbols) and of the 573 pre-selected stars having $\varpi/\delta \varpi>10$, $r_{apo}>75\kpc$ and $d_\odot < 4\kpc$ (denoted by circles). Stars are colored by their apocenter values in the upper panel and and by their vertical action values in the lower row of panels.
   Upper panel: $(J_\phi , J_z)$ plane used for the selection where the overdensity was discovered. The most significant detection obtained using a Hough transform technique \citep{Illingworth1988} on stars with $J_z > 1000 \kpc \kms$ (i.e. with large departures from the Galactic mid-plane) is shown with a red line. This line runs through the Typhon structure. The parallelogram selection of the structure is depicted in a solid line encompassing 16 stars, and is defined by: $J_z \in [2000, 3100] \kpc \kms$ and $3.3 J_\phi + 3500 \kpc \kms < J_z < 3.3 J_\phi + 5000 \kpc \kms$. The symmetric (retrograde) selection with respect to the $J_\phi=0$ line is shown with a dashed line.
   Bottom-left and bottom-right panels respectively show the $(J_\phi , J_r)$ and $(J_\phi , E_{tot})$ planes.}
\label{fig:selection}
\end{figure*}

The Gaia mission has recently made accessible its third data release (DR3) \citep{GaiaDR3} of its all-sky survey. It contains approximately 33 million stars with mean radial velocities down to $G \sim 15$, which, complemented with the excellent proper motions and parallaxes published in the earlier EDR3 release \citep{2021A&A...649A...1G}, provide the required phase-space constraints. Because the DR3 radial velocity limit is quite shallow, it almost exclusively probes the very nearby regions of the Galaxy (the median distance of the sample with $10\sigma$ parallaxes is only $1.26\kpc$). In the vicinity of the Sun, dynamical times are short and tidal debris are expected to phase-mix rapidly \citep{1999Natur.402...53H,2003MNRAS.339..834H}, erasing any initial stream-like coherence that might have been present. 

\begin{figure*}[!htb]
\centering
    \includegraphics[angle=0,  clip, width=\textwidth]{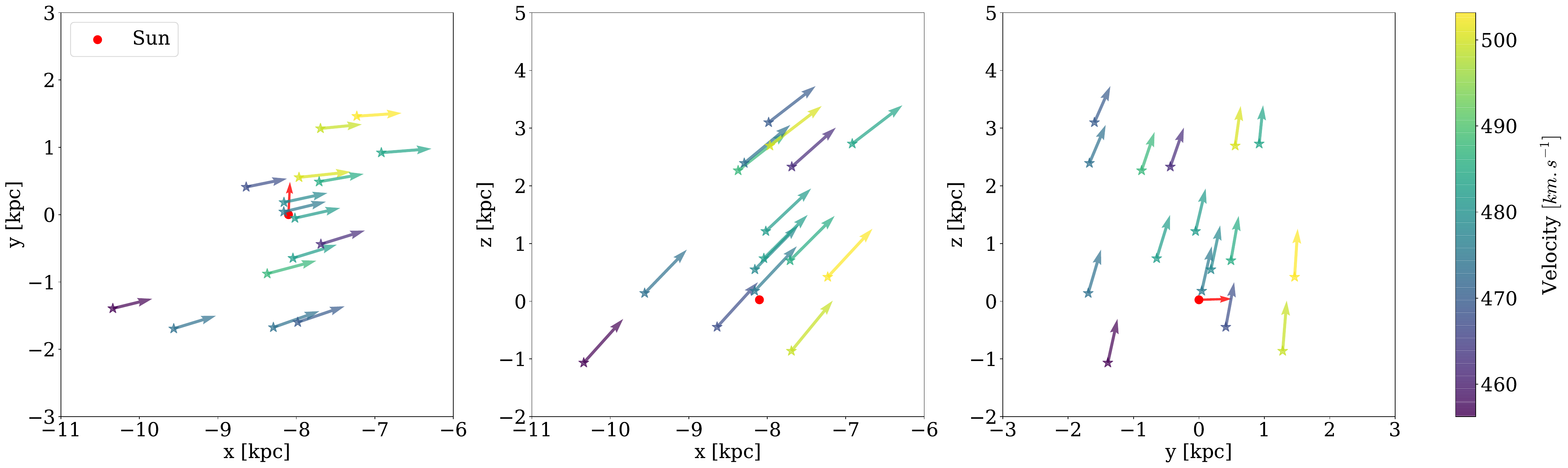}
   \caption{
   Positions and velocity vectors in galactic Cartesian coordinates of Typhon sample members. Velocity vectors scale $1:3\times 10^3$. The sample shows very clear streaming motion. For reference, the position and velocity vector of the Sun is also shown in red.}
\label{fig:posvel}
\end{figure*}

In this contribution we show that, surprisingly and contrary to those expectations, the Solar neighborhood contains a very wide yet kinematically coherent metal poor stellar stream, which we name Typhon\footnote{The serpent Typhon is the child of Gaia and Tartarus (the deep abyss) in Greek myth.}, whose apocenter reaches out to $> 100\kpc$ -- the edge of the Galactic halo.


\section{Selection}
\label{sec:selection}


From the Gaia DR3 catalogue, we select the 25,355,580 stars with well-constrained distances ($\varpi/\delta \varpi>10$), radial velocities measured by Gaia's Radial Velocity Spectrometer (RVS) instrument \citep{2022arXiv220605541R}, having at least a 5-parameter astrometric solution, and with magnitudes in the range $0\le G \le 22$, $0\le G_{BP} \le 30$, $0\le G_{RP} \le 30$. To convert the apparent motions to motions in a frame\footnote{Throughout the paper, we use a right-handed Galactic Cartesian coordinate system.} at rest with respect to the Galaxy, we assume that the Sun is located at $(x,y,z)_\odot=(-8.2240, 0, 0.0028) \kpc$ (Solar radius from \citealt{2020arXiv201202169B} and $z$-position of the Sun from \citealt{2021A&A...646A..67W}), and that it moves with a peculiar velocity $(v_x, v_y, v_z)_\odot=(11.10, 7.20, 7.25)\kms$ (\citealt{2010MNRAS.403.1829S}, with the $\phi$-direction velocity from \citealt{2020arXiv201202169B}), and we take the circular velocity at the Solar radius to be $243\kms$ \citep{2020arXiv201202169B}. We use the resulting phase space measurements to derive the orbital parameters of the stars, including the pericenter and apocenter distances, as well as action-angle coordinates calculated using the AGAMA package \citep{Vasiliev2019} in a realistic potential model \citep{McMillan2017} for the Milky Way. Since we are particularly interested in finding debris from the outer halo that could be associated to ancient merger events, we impose an apocenter cut at $r_{apo}>75\kpc$, which yields a sub-sample of 870 stars.\\


Further analysis is performed in the space of actions $(J_r,J_\phi , J_z)$, which encode, respectively, the amplitude of orbital motion in the radial, azimuthal, and vertical directions. In particular, we plot the $(J_\phi , J_z)$ projection colored by $r_{apo}$ in Figure \ref{fig:selection}. There, a polar structure can be spotted as a tight, almost vertical, linear grouping between $(J_\phi \sim -650 \kpc \kms, J_z \sim 2100 \kpc \kms)$ and $(J_\phi \sim -400 \kpc \kms, J_z \sim 3000 \kpc \kms)$. We find that this feature is most striking when the sample is limited to stars with heliocentric distances $d_\odot < 4\kpc$, approximately at the limit of useful 6-D phase-space data in the DR3 catalog. In particular, performing the Hough transformation \citep{Illingworth1988} line detection technique on the stars in the $(J_\phi , J_z)$ plane (binning the action data into pixels of size $30 \kpc \kms$ on a side and adopting a $1 \deg$  discretization for the angle of the fitted lines), we find that the most significant linear grouping of stars with $J_z > 1000 \kpc \kms$ (i.e. that experience large excursions from the Galactic mid-plane) corresponds to this quasi-linear overdensity (red line in Figure \ref{fig:selection}). These 16 stars possess similar apocenter distances ($r_{apo} \approx 100\kpc$), and are also highly correlated in the angle coordinates $(\theta_r,\theta_\phi,\theta_z)$ conjugate to the actions.

We then separate this structure from the bulk of the data by applying a simple parallelogram selection in the $(J_\phi, J_z)$ plane, as follows: $J_z \in [2000, 3100]$ and $3.3 J_\phi + 3500 < J_z < 3.3 J_\phi + 5000$, which results in a final sample of 16 stars. This selection box is displayed as a solid black parallelogram in Figure~\ref{fig:selection}.\\


Furthermore, it should be noted that the symmetric control selection around $J_{\phi}=0$, shown as a dashed parallelogram, encompasses only two stars and they do not possess homogeneous dynamical properties. Assuming that the halo is symmetric in angular momentum, there is no a priori reason for the prograde selection to contain significantly more stars than the symmetric retrograde selection as is the case here, other than the selection containing a coherent dynamical group. Taking the symmetric selection as a control sample we estimate the significance of the detection to be $\approx 3.5\sigma$. We note in passing that the Gaia Universe Model Snapshot (GUMS, \citealt{Gums}, updated for Gaia DR3) contains no artificial stars with the selection criteria used to detect Typhon, suggesting that Typhon is a coherent structure that can only be explained by an external body not included in that simulation.


\section{Characteristics}
\label{sec:characteristics}


The positions and velocities of the sample members of the Typhon stream are shown in Figure \ref{fig:posvel}. We find that member stars of this polar stream are spread out all around us, passing through the Solar neighborhood with a high vertical velocity, and exiting the disk at an angle of $\sim 50\degree$ with respect to it.

\begin{figure*}[!htb]
\centering
    \includegraphics[angle=0,  clip, width=\textwidth]{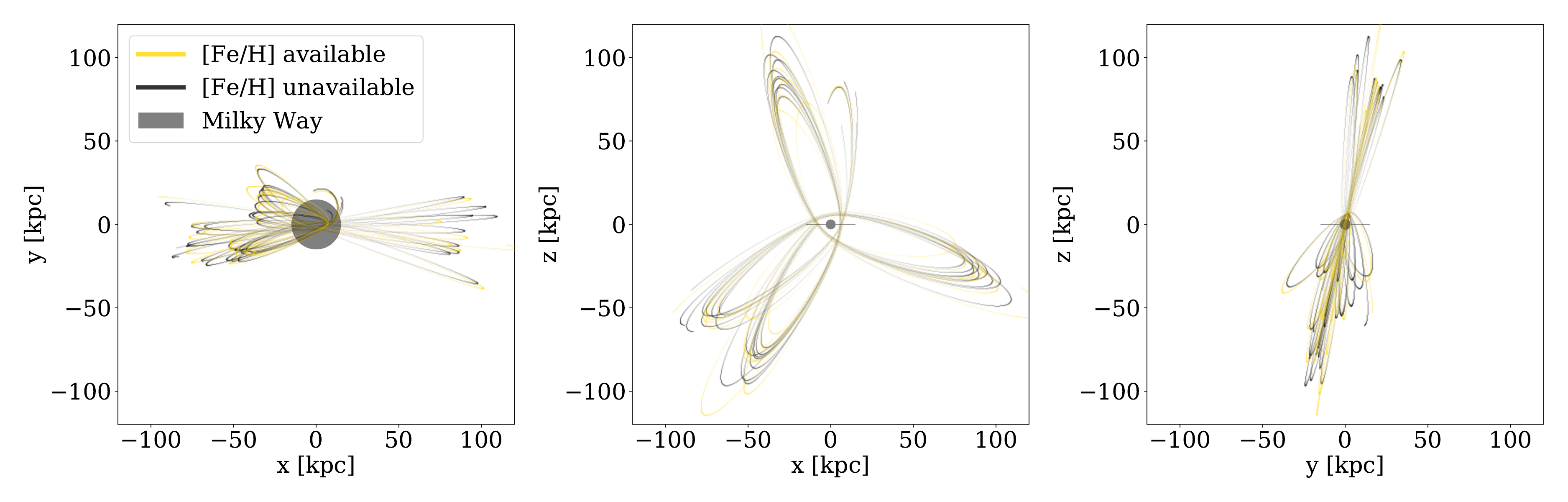}
   \caption{Trajectories of the sample members of Typhon during a $5\Gyr$ backward integration in the \citep{McMillan2017} potential in galactic Cartesian coordinates. Trajectories of the 7 stars whose metallicity is available through LAMOST DR8 \citep{LAMOSTDR8} are colored in yellow.}
\label{fig:orbits}
\end{figure*}

In Figure \ref{fig:orbits} we show the result of integrating Typhon members backwards in time for $5\Gyr$ in the \citep{McMillan2017} Milky Way potential model. Although the stars were selected from a small region in the $(J_\phi, J_z)$ plane (but with no constraint on $J_r$), and so should therefore possess similar orbits, there was no a-priori reason for the sample to be in phase, as is clearly the case from an inspection of Figure \ref{fig:orbits}. The sample is dynamically coherent, with very similar orbital parameters: $r_{peri}=6.0\pm0.5\kpc$, $r_{apo}=99\pm15\kpc$, $J_r=6400\pm1000 \kpc \kms $, $J_\phi=-560\pm110 \kpc \kms $, $J_z=2500\pm300 \kpc \kms$ and eccentricity $e=0.88\pm0.02$.

We estimate the 3-dimensional velocity dispersion of the stream to be $\sigma_{v, 3D} \approx 13 \kms$ by considering the velocity differences of the stars to the computed orbit of the star with {\it Gaia} ID 3939346894405032576 (whose orbit through the Solar neighborhood appears closest to the middle of the sample). Assuming isotropy, the one-dimensional velocity dispersion is then $\sigma_{v} \approx 7.5 \kms$.\\

We cross-matched our sample with the LAMOST DR8 \citep{LAMOSTDR8} catalog, in particular the ``FEH\_PASTEL'' column which covers a wide range of metallicities especially on the very metal-poor regime, enabling us to obtain high quality spectroscopic metallicities for 7 stars
of the Typhon stream (the stellar parameters of which lie within the reliable range of the PASTEL catalog). These measurements span between ${\rm [Fe/H]}=-2.23\pm0.06$~dex and ${\rm [Fe/H]}=-1.25\pm0.09$~dex.

\begin{figure*}[!htb]
\centering
\hbox{
    \includegraphics[angle=0,  clip, width=0.5\textwidth]{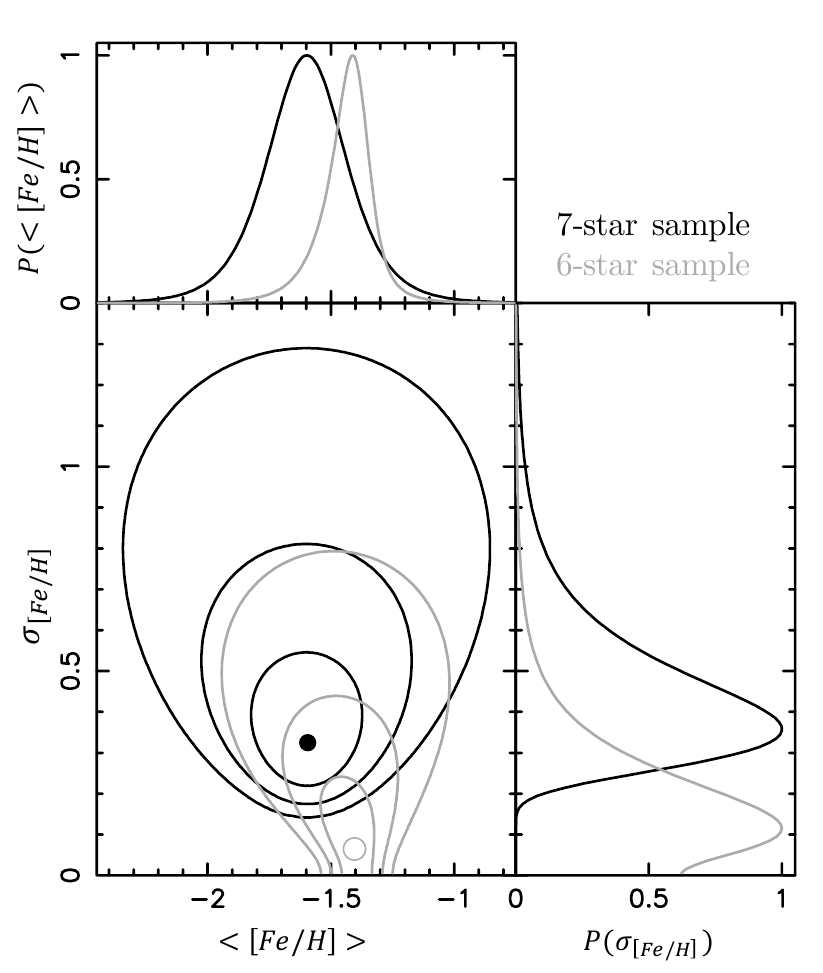}
    \includegraphics[angle=0,  clip, width=0.5\textwidth]{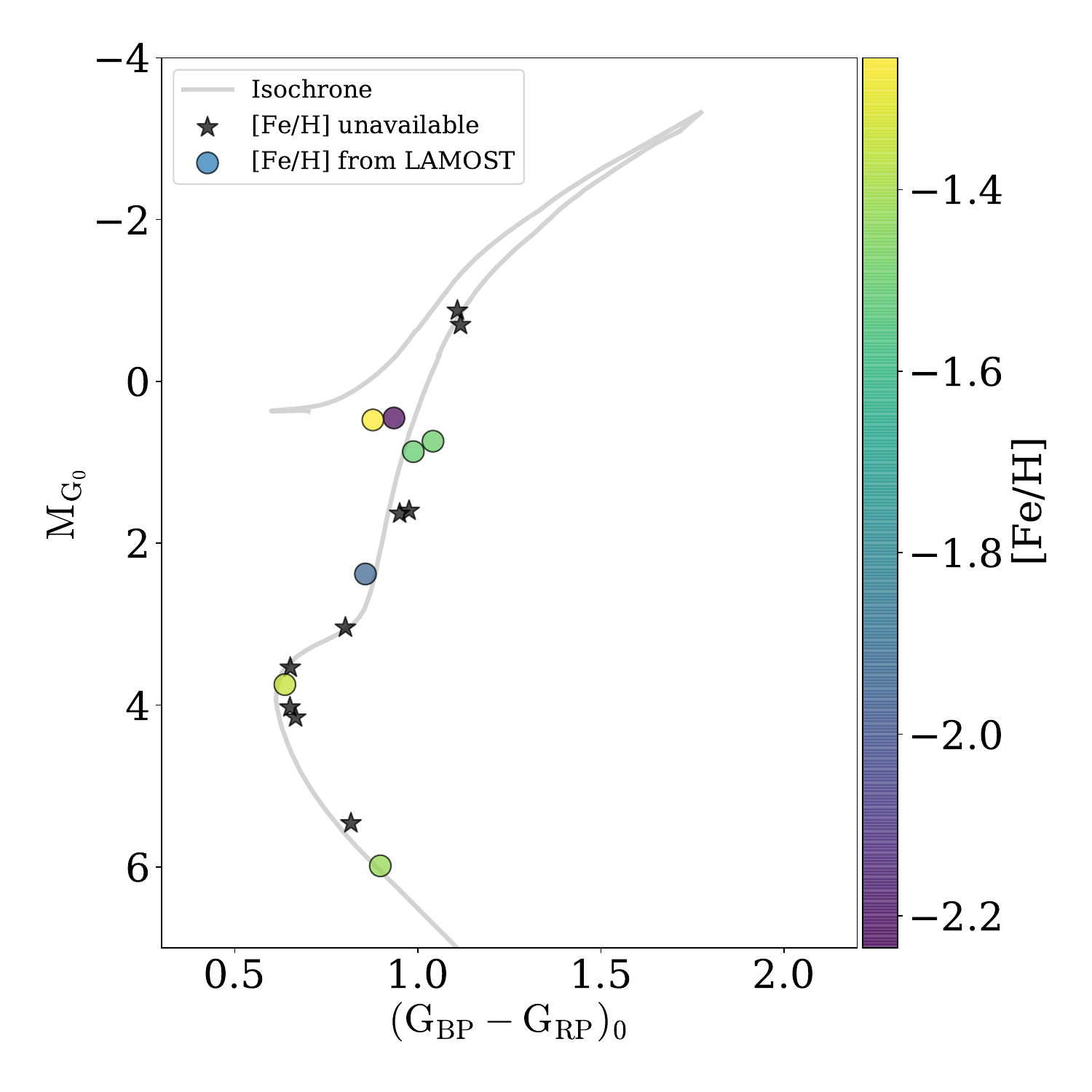}}
    \caption{Left: Likelihood contours of the mean metallicity and metallicity dispersion of the spectroscopic sample, shown for the full 7 star sample (black lines), and removing the most metal poor star (grey lines). Right: Color magnitude diagram of the sample members of Typhon. For reference, the grey line shows a PARSEC isochrone model \citep{2012MNRAS.427..127B} of age $12.5\Gyr$ and of metallicity ${\rm [Fe/H]} = -1.60$~dex. The reasonable correspondence of this model shows that the population is predominantly very old.}
\label{fig:CMD}
\end{figure*}

As shown in Figure~\ref{fig:orbits}, where we color orbits of stars of known metallicity in yellow, these stars are dynamically representative of the full sample. 
In Figure~\ref{fig:CMD} (left panel), we show the likelihood distribution (black contour lines) for the mean metallicity and for the intrinsic dispersion of the metallicity distribution (correcting for the LAMOST uncertainty estimates, assuming that they are reliable). We find $\langle {\rm [Fe/H]}\rangle = -1.60^{+0.15}_{-0.16}$~dex, and $\sigma({\rm [Fe/H]}) = 0.32^{+0.17}_{-0.06}$~dex, which indicates that the system has a resolved dispersion in metallicity. We note however, that this result depends on the inclusion of the most metal-poor star in the sample; if it is removed (although we have no a-priori reason to do so) these values become $\langle {\rm [Fe/H]}\rangle = -1.41^{+0.05}_{-0.09}$~dex, and $\sigma({\rm [Fe/H]}) = 0.06^{+0.17}_{-0.06}$~dex, consistent with no dispersion at the $1\sigma$ level.

These metallicities are consistent with the color magnitude diagram shown in Figure~\ref{fig:CMD}, where we use the 3D extinction estimates by \citealt{2022A&A...658A..91A} to deredden the stars. In addition, based on the PARSEC stellar population models \citep{2012MNRAS.427..127B}, and using the canonical two-part-power law initial mass function corrected for unresolved binaries \citep{Kroupa2021}, and Gaia's detection limit, we compute the order of magnitude of the density of the Typhon stream to be of $\sim 25 \msun / \kpc^{-3}$ in the $d_\odot < 1.5 \kpc$ solar vicinity fragment. However, without further information we refrain from extrapolating this value out to compute the mass of the full stream structure.


\section{Discussion and conclusions}
\label{sec:discussion}

Although the search for new stellar streams is currently a very active field, to the best of our knowledge the structure discussed here (Typhon) that we isolated thanks to the new and excellent Gaia DR3 data was never identified before. It should be noted that although Typhon is very close to the DTG-11 stream identified in \citep{Yuan2020} in the $(J_\phi , J_z)$ plane, we verified that Typhon is a distinct structure. In particular, we see that Typhon members have much higher apocenters ($\approx 100\kpc$ vs. $\approx 15\kpc$ for DTG-11) which becomes obvious when comparing their very different $J_r$ values. In addition, we compared our sample to the thorough \citet{Malhan2022} atlas of stellar streams and found no previously mapped equivalent structure. We note that the discovery of the Typhon structure was confirmed by \citet{Dodd2022} shortly after the first submission of our paper using a formal clustering metric.\\

In addition, a follow up study focusing on the chemical abundances of Typhon was published by \cite{Ji2022}. That contribution presents high resolution spectra for 7 Typhon members chosen solely based on observability, including 3 members whose metallicities are not available in LAMOST DR8, which nevertheless show consistent metallicities with the  LAMOST sub-sample, thereby supporting our conclusions regarding the metallicity distribution of the structure.

The characteristics of Typhon members given in Section \ref{sec:characteristics} lead us to believe that Typhon is likely the tidal remnant of a dwarf galaxy. In particular the metallicity spread, vertical action spread and structure width appear completely incompatible with a globular cluster progenitor. With metallicities reaching ${\rm [Fe/H]} \sim -1.3$~dex, and with a mean of ${\rm [Fe/H]} \sim -1.6$~dex, the
mass-metallicity relation of dwarf galaxies \citep{Kirby2013} suggests that the progenitor likely possessed a luminosity of $10^6$ -- $10^7 {\rm \, L_\odot}$, perhaps similar to the Sculptor dSph. \cite{Ji2022} concur with us on this point. The estimated velocity dispersion value of $\sigma_{v} \approx 7.5 \kms$ lies between that of the Orphan Stream ($\sigma_{v} \approx 5 \kms$, \citealt{2019MNRAS.485.4726K}) and the stream of the Sagittarius dwarf galaxy ($\sigma_{v} \approx 13 \kms$, \citealt{2017MNRAS.464..794G}), suggesting that the mass of the Typhon progenitor likely exceeded $10^8\msun$ (an estimate for the mass of Orphan Stream progenitor, \citealt{2019MNRAS.486..936F}), but was not as massive as the Sagittarius dwarf.


We noticed that although in the heavy \citet{McMillan2017} gravitational potential $(M_{\rm vir}=1.3\times10^{12}\msun)$ all stars in the sample are bound, in the lighter MWPotential2014 (\citealt{Bovy2015}, $M_{\rm vir}=8\times10^{11}\msun$), half of the Typhon stream members are unbound\footnote{Note that none of the Typhon stars were flagged as hyper-velocity stars in the \cite{Marchetti2018} census.}. This underlines how having constraints on the trajectories of streams such as Typhon is of great value as the trajectories of these streams are very dependant on the acceleration field of the Milky Way and its underlying dark matter distribution. 

We also checked whether the Typhon members could have close encounters with the Large Magellanic Cloud (LMC) or the Sagittarius dwarf galaxy. Taking the trajectories of the two satellites from \citet{2021MNRAS.501.2279V}, we find that the LMC remains always very distant ($\simgt 40\kpc$). However, the Typhon stars probably did experience a relatively close flyby of Sagittarius ($\sim 20\kpc$, $0.10\Gyr$ ago). We note that Typhon and Sagittarius share very similar orbital planes, although they possess opposite angular momentum vectors (i.e. the direction of motion in the plane is opposite). The interaction between Typhon and Sagittarius will be interesting to analyse with $N$-body simulations, but we defer that investigation to a future contribution.\\


The identification of this high apocenter polar stream passing so close to the Sun raises many questions. Assuming that the Solar vicinity is not special and is representative of an average location in the disk, the present detection could be used to place constraints on the number of highly radial accretions that took place during the formation of the Milky Way. The picture suggested by Typhon is that there may be a large population of outer halo dwarf galaxies or dwarf galaxy fragments residing near their apocenters, akin to the ``Oort Cloud'' around the Sun. A more thorough survey of local phase space for other Typhon-like structures and also deeper next-generation sky surveys (with LSST, for instance) that might detect them in place in the outer halo will help quantify this possibility.

This discovery also underlines the relevance of stream research in the Solar vicinity where great quantities of high quality data are available in addition to spatially wider searches. This poses several challenges and may require the development of new algorithmic approaches suited to exploit \textit{Gaia} era data for nearby structures with incomplete astrometry (e.g. missing line of sight velocities) as sections of streams passing near us are not easily identifiable as streams when projected onto a sky map.

In future work, it will be very useful to attempt to extend the detections along the stream so as to chart it out further in its orbit through the Galaxy. As we alluded to above, such stars may provide very useful dynamical probes for the Milky Way's dark halo, and they will be invaluable to inform follow-up simulation studies attempting to model the N-body evolution of the system. Similarly, having full metallicity information for the member stars would be of great value in order to confirm the present hypothesis regarding the nature of the progenitor.\\

\section*{Data Availability}
\label{sec:Data}

The final sample is provided in Table \ref{tab:data} with a more complete table available at \href{https://doi.org/10.5281/zenodo.6979887}{DOI 10.5281/zenodo.6979887}.

\begin{deluxetable*}{lcccccccccccc}
\centering
\tablehead{
    \colhead{\textit{Gaia} source ID} & \colhead{RA} &
    \colhead{DEC} & \colhead{$J_r$} & \colhead{$J_z$} & \colhead{$J_\phi$} & \colhead{$r_{\rm peri}$} & \colhead{$r_{\rm apo}$} & \colhead{${\rm FeH}$}\\
    \colhead{} & \colhead{[deg]} &
    \colhead{[deg]} & \colhead{$[\kpc\kms]$} & \colhead{$[\kpc\kms]$} & \colhead{$[\kpc\kms]$} & \colhead{$[\kpc]$} & \colhead{$[\kpc]$} & \colhead{[dex]}
    }
    \tablewidth{\textwidth}
    \tablecaption{The Typhon sample. Orbital parameters derived using a \citep{McMillan2017} potential. Metallicities from LAMOST DR8 (PASTEL column) \citep{LAMOSTDR8} are listed when available. \label{tab:data}}
    \startdata
    291927350856672768   &    23.65   &    24.43   &    5467.78  &    2109.67  &    -678.85  &    5.64    &    85.13   &          -            \\
    1255095276181144320  &    218.71  &    25.17   &    5500.85  &    2153.71  &    -641.67  &    5.65    &    85.59   &  $-1.42  \pm   0.11$  \\
    1264504793612855808  &    226.50  &    24.27   &    5880.03  &    2536.48  &    -561.03  &    6.11    &    92.86   &  $-1.50  \pm   0.41$  \\
    1303595799235740288  &    243.67  &    25.97   &    5957.51  &    2210.55  &    -625.38  &    5.63    &    91.92   &          -            \\
    1470463353223683840  &    205.82  &    33.33   &    8919.22  &    2929.18  &    -555.25  &    6.65    &    138.32  &  $-1.94  \pm   0.22$  \\
    1485076859188949760  &    212.45  &    37.57   &    5789.44  &    2253.55  &    -641.30  &    5.81    &    90.15   &          -            \\
    1765600930139450752  &    327.45  &    10.81   &    7675.31  &    2069.85  &    -643.15  &    5.17    &    114.35  &  $-2.24  \pm   0.06$  \\
    3013164238138435712  &    82.50   &    -10.56  &    6320.46  &    2733.22  &    -343.02  &    6.15    &    98.75   &          -            \\
    3029220853122930432  &    115.02  &    -14.82  &    6971.03  &    2793.03  &    -391.92  &    6.34    &    108.47  &          -            \\
    3573787693673899520  &    181.02  &    -13.37  &    6106.32  &    2288.46  &    -607.18  &    5.75    &    94.38   &          -            \\
    3736372993468775424  &    197.96  &    11.29   &    4584.30  &    2469.84  &    -651.10  &    6.44    &    76.04   &  $-1.50  \pm   0.09$  \\
    3793377208170393984  &    173.49  &    -2.47   &    6185.63  &    2884.76  &    -441.23  &    6.65    &    98.78   &  $-1.25  \pm   0.09$  \\
    3891712266823336192  &    181.86  &    1.68    &    5794.96  &    3007.81  &    -400.13  &    6.87    &    94.16   &          -            \\
    3913243629368310912  &    178.32  &    10.63   &    7589.72  &    2826.15  &    -502.54  &    6.47    &    118.11  &          -            \\
    3939346894405032576  &    200.01  &    19.69   &    6128.64  &    2412.30  &    -602.82  &    5.99    &    95.58   &  $-1.35  \pm   0.07$  \\
    4537774136693362944  &    280.74  &    26.22   &    7212.08  &    2163.32  &    -670.81  &    5.49    &    108.86  &          -            \\
    \enddata
\end{deluxetable*}

\begin{acknowledgments}
The authors thank the anonymous referee for their very helpful comments and acknowledge funding from the European Research Council (ERC) under the European Unions Horizon 2020 research and innovation programme (grant agreement No. 834148) and from the Agence Nationale de la Recherche (ANR projects ANR-18-CE31-0006, ANR-18-CE31-0017 and ANR-19-CE31-0017).
This work has made use of data from the European Space Agency (ESA) mission {\it Gaia} (\url{https://www.cosmos.esa.int/gaia}), processed by the {\it Gaia} Data Processing and Analysis Consortium (DPAC, \url{https://www.cosmos.esa.int/web/gaia/dpac/consortium}). Funding for the DPAC has been provided by national institutions, in particular the institutions participating in the {\it Gaia} Multilateral Agreement.

\end{acknowledgments}

\break

\bibliography{halo_alien}
\bibliographystyle{aasjournal}

\end{document}